\documentclass[journal]{IEEEtran}

\usepackage{soul} 
\usepackage{color, xcolor} 
\usepackage{amsmath,amsfonts}
\usepackage{algorithmic}
\usepackage{array}
\usepackage{textcomp}
\usepackage{stfloats}
\usepackage{url}
\usepackage{verbatim}
\usepackage{graphicx}
\usepackage{amssymb}
\usepackage{color}
\usepackage{multirow}
\usepackage{longtable}
\usepackage{amsmath,amsthm,amssymb,amsfonts}
\usepackage{graphicx}
\usepackage{epstopdf}
\usepackage{epsfig}
\usepackage{marvosym}
\usepackage{subcaption}

\usepackage{longtable}
\usepackage{rotating}

\def\BibTeX{{\rm B\kern-.05em{\sc i\kern-.025em b}\kern-.08em
    T\kern-.1667em\lower.7ex\hbox{E}\kern-.125emX}}
\usepackage{balance}
\begin{document}
\title{A graph neural network based on feature 
network for identifying influential nodes}
\author{Yanmei Hu,Siyuan Yin,Yihang Wu,Xue Yue and Yue Liu
\thanks{Manuscript created january, 2024. \textit{(Corresponding author: Yanmei Hu.)}
\par
Yihang Wu,Siyuan Yin,Xue Yue and Yue Liu are with College of Computer and Cyber Security, Chengdu University of Technology, ChengDu 610000, China.
\par
Yanmei Hu is with College of Computer and Cyber Security,Chengdu University of Technology, ChengDu 610000, China (e-mail: huym260@126.com).
}}

\markboth{Journal of \LaTeX\ Class Files,~Vol.~18, No.~9, September~2020}%
{How to Use the IEEEtran \LaTeX \ Templates}

\maketitle

\begin{abstract}
  Identifying influential nodes in  complex networks is of great importance, and has many applications in practice. For example, finding  influential nodes in e-commerce network can provide merchants with customers with strong purchase intent; identifying influential nodes in computer information system can help locating the components that cause the system break down and identifying influential nodes in these networks can accelerate the flow of information in  networks.  Thus, a lot of efforts have been made on  the problem of indentifying influential nodes. However, previous efforts either consider only one aspect of the network structure, or using  global centralities with high time consuming as node features to identify influential nodes, and  the existing methods do not consider the relationships between different centralities. To solve these problems, we propose a Graph Convolutional Network Framework based on Feature Network, abbreviated as FNGCN (graph convolutional network is abbreviated as GCN in the following text). Further, to exclude noises and reduce redundency, FNGCN utilizes feature network to represent the complicated relationships among the local centralities, based on which the most suitable local centralities are determined. By taking a shallow GCN and a deep GCN into the FNGCN framework, two  FNGCNs are developed. With ground truth obtained from the widely used Susceptible Infected Recovered (SIR) model, the two FNGCNs are compared with the state-of-art methods on several real-world networks. Experimental results show that the two FNGCNs can identify the influential nodes more accurately than the compared methods, indicating that the proposed framework is effective in identifying influential nodes in complex networks.
\end{abstract}


\section{Introduction}
\IEEEPARstart{T}{he} rapid growth of the Internet and the exponential growth of data  lead to increasingly complex network structures, many researchers  pay more and more attention to the study of complex networks composed of various information. How to identify influential nodes in complex networks is a research direction that is gradually attracting attention, which can be applied to recommender systems   \cite{2023R}, fault diagnosis \cite{2008A}, object detection \cite{2022O}, and social networks \cite{2022SN}. For example, influential nodes are able to interact with other nodes with more information, which improves the overall efficiency of the recommendation system; identifying influential nodes can also provide  better nodes for informatiozn propagation and accelerate social information flow in social networks. 

\noindent There have been many ways to identify influential nodes in the literature \cite{2020R,2020inf,2022cgcn}. The traditional methods  determine the influence of each node by scoring it through a metric directly based on the network structure. The examples are node centrality-based methods \cite{2005C} and K-shell algorithm \cite{2007K}. Because these methods are based on a single metric that considers only one aspect of the network structure, e.g., the degree centrality of a node only counts its neighbors, the betweenness centrality only considers whether a node is on the shortest path of many pairs of node; a large deviation may be led in the result of those methods, making them not suitable for most scenarios. Machine learning provides a new research direction for identifying influential nodes of the network, producing several machine learning-based methods to identify influential nodes. The machine learning-based methods construct node features from the network structure (e.g., taking node centralities as node features) and feed node features to  a specific model, e.g., Support Vector Machines (SVM) \cite{1998S}, Logistic Regression (LR) \cite{2008L} and RCNN \cite{2020R} to obtain influential nodes.  With the introduction of   Graph Neural Networks (GNNs) \cite{2020G}, the GNN-based methods input  the  network structure and  feature matrix   into a multi-layer neural network to obtain the influential nodes, and the network structure  is usually represented as a normalized laplacian matrix \cite{2020inf} or a transformed transition matrix \cite{2020GCNII}.  For example, InfGCN\cite{2020inf} combines local centralities (degree and clustering coefficient) and global centralities (betweenness and closeness) to obtain node features, and then input these node features and the normalized laplacian into a  Graph Convolutional  Network (GCN) \cite{2016GCN} model. Compared with machine learning-based methods, GNN-based methods input not only the node features but also the network structure. Because of the addition of the network structure as input and the multiple nonlinear transformation of multi-layer neural networks, these methods are more effective in identifying influential nodes. However, the high time consumption of computing global centralities reduces the efficiency of these methods. Moreover, the existing GNN-based methods do not consider the relationships between different centralities, leading to feature redundancy; and most of the GNN-based methods use shallow models, without exploring the effectiveness of deep models on identifying influential nodes.
\par
\noindent To solve these problems above, we propose a Graph Convolutional Network framework based on Feature Network (FNGCN), which only considers local centralities and utilizes a feature network constructed from those local centralities to prepare node features. This framework first constructs a feature network to represent the complicated relationships among different local centralities, based on which the node features is obtained. After that, the adjusted transition matrix and the obtained feature matrix (formed by node features) are fed into a GCN model. Specifically, a shallow GCN and a deep GCN are applied to the FNGCN framework, producing two FNGCNs. Finally, the two FNGCNs are trained on the labelled data from the Susceptible Infected Recovered (SIR) model\cite{2009SIR}. To test the proposed FNGCN, five experiments are conducted on several real-world networks. The first experiment is to compare FNGCN with other methods, and the last four experiments are ablation experiments to explore the following four questions, which are: 1) what's the influence of the number of hidden GCN layers on FNGCN; 2) are local centralities sufficient for GCN to identify influential nodes; 3) is it necessary to select appropriate local centralities; 4) what's the contribution of each node feature (the selected local centrality) to GCN-based models. The experimental results show that the two FNGCNs perform better on identifying influential nodes than the compared methods, and the FNGCN framework consumes much less time on feature construction than other deep learning methods, since it only considers the local centralities to prepare node features. Moreover, although deep FNGCNs perform better than shallow FNGCN overall, the later ones is not much inferior to the former ones; local centralities are sufficient for GCN to identify influential nodes; selecting appropriate local centralities is necessary, and the centrality of Conductance of Egonet is the most important feature to FNGCN. The main contributions of this paper are as follows:

\noindent 1) We propose FNGCN, a framework based on feature network for the identification of influential nodes in complex networks. Unlike the existing GNN-based methods that directly take several centralities as node features, FNGCN constructs a feature network to explore and analyse the complex relationships between different local centralities, and selects the most appropriate centralities as node features. After that, FNGCN applies a GCN to distinguish influential nodes according to network structure and the obtained node features.

\noindent 2) We explore the influence of the number of hidden GCN layers on FNGCN, the ability of local centralities for GCN to identify influential nodes, the necessity to select centralities as node features, and the contribution of each feature (a centrality) to the identification model.

\noindent 3) We conduct comprehensive experiments across several datasets to demonstrate that FNGCN is superior to the state-of-the-art methods. Moreover, deep FNGCN is marginally superior to shallow FNGCN, local centralities are sufficient to construct node features for GCN, but selecting appropriate centralities is necessary, the centrality of Conductance of Egonet is most important to FNGCN.
\noindent The rest of the paper is organized as follows. In Section 2, we present related work. Section 3 describes our proposed method in details. Section 4 conducts extensive experiments to test FNGCN. Finally, Section 5 concludes our work.

\section{Related Work}
\noindent Effectively identifying influential nodes in complex social networks is a challenging problem. Researchers propose many methods in this field, which can be roughly divided into traditional methods, machine learning-based methods and GNN-based methods. 

\noindent\textbf{Traditional methods}. The node centrality of the network is a classic method for identifying influential nodes, and it is also an important centrality that we need to use in our experiments. There are many classes of node centrality, which can be roughly divided into global centrality and local centrality.  However, existing centrality-based methods for identifying influential nodes are basically based on global centrality and do not use local centrality to identify influential nodes in the network.  Each class of global centrality can be subdivided according to its definition. For example, the closeness centrality \cite{2008clo} and betweenness centrality \cite{2004be}  are measured by the shortest path between nodes, while PageRank \cite{2011pg} and eigenvector centralities \cite{2000eg}  evaluating  nodes mainly based on the importance of nodes' neighbors to obtain node centrality.  K-shell iteratively deletes nodes whose degree  is less than or equal to $k$ until there are no nodes with degree $k$ in the network. In order to solve the problem that the K-shell algorithm has many nodes with the same $k$ value in each-shell, researchers add entropy to nodes with the same $k$ value to distinguish the influence of nodes.

\noindent \textbf{Machine learning-based methods and GNN-based methods}.  The machine learning-based methods construct node features from the network structure and feed node features to a specific model. Some researchers combine local centrality and global centrality to obtain node  features, and then input node features into Support Vector Machines (SVM) and Logistic Regression (LR) to  obtain the influential nodes \cite{2020inf}. The  RCNN \cite{2020R} adopt  Convolutional Neural Networks (CNNs) \cite{2019G}   to extract node features from the subgraph of each node and output a score of each node, then feeding  the score and label of each node  to the squared loss function to train the model, finally the score of each node in the test dataset are obtained without label using the trained RCNN model. Compared with traditional methods, the machine learning-based methods introduce machine learning models such as SVM, LR and RCNN based on multiple centralities, enhancing their ability to identify influential nodes.

\noindent GNNs provides a new research direction for identifying important nodes in the network by combining the network structure  and node features. The GNN-based methods take the network structure and node features as input of a multi-layer neural network to obtain the influential nodes. Thomas N. Kipf and Max Welling propose a new semi-supervised model baesd on GCN for node classification\cite{2016GCN} . Gouheng Zhao, et al. first construct a normalized laplacian matrix of the network, then input the normalized laplacian matrix to InfGCN model along with the node features\cite{2020inf} . In addition, Min Zhang, et al. propose CGCN  by combining CNN and GCN\cite{2022cgcn}  CGCN first uses the contraction algorithm  to extract the node features from the subgraph of each node, and then inputs node features and normalized laplacian matrix into the CGCN model composed of CNN and GCN to obtain the influence score of each node. Chen et al. proposed a GCNII which is a deep GCN to solve the problem that most of the current GNN-baesd models have a shallow architecture that limits their model performance due to over-smoothing \cite{2020GCNII}. To aviod the over-smoothing, GCNII takes two techniques, which are initial residual and identity mapping. Our work falls into the line of GNN-based methods to identify influential nodes, but different from the previous works, we only use local centralities which reduces time consumption of constructing node features; moreover, some existing works in other domains such as recommendation and cancer diagnosis have shown the importance to explore the relationships between different features \cite{2021FG-RS,2023fl}, thus we consider the relationships between diffrent centralities, and select proper local centralities from the view of network to obtain node features. In addition, on one hand, deep models attract more and more attention because their strong learning ability in diverse domains\cite{2022mc}, it is meaningful to explore the influence of the number of hidden GCN layers on the final model; on the other hand, decoupling different features/factors \cite{2023nf} to explore their importance/contributions to the model can improve the interpretability of the model; thus we further explore the influence of the number of hidden GCN layers and the contributions of each node feature to the proposed model.

\section{Method}
\noindent  In this section, we design an effective graph neural network framework based on feature network, aiming at identifying influential nodes with only network structure. It is known that constructing proper centralities for nodes is essential to GNNs \cite{2019importanceoffeature}. Although many centralities, such as degree, local clustering coefficient and PageRank score, can be obtained from network structure, not all centralities are proper, and some centralities may be redundant and take nosies. In order to construct proper centralities, we explore the centrality space from the perspective of network, and based on the exploration, we design an effective graph neural network named as FNGCN. Next, we first present the framework of FNGCN, and then describe centrality construction on details, and finally present the SIR model used to label influential nodes.

\subsection{Framework} \label{subsec:Framework}

\begin{figure}[ht]
  \centering 
  \includegraphics[height=4.1cm,width=8cm]{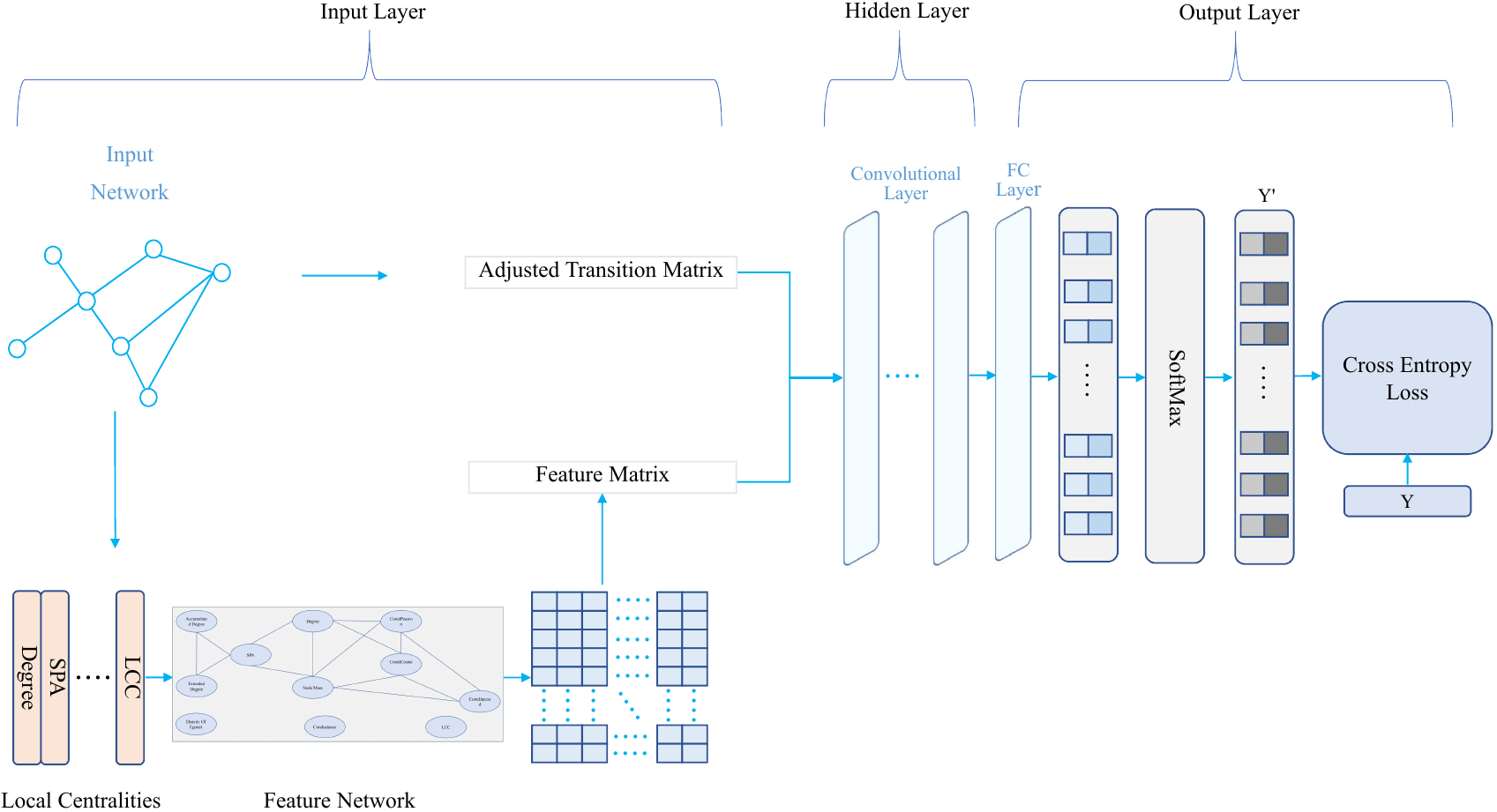}
  \caption{The framework of FNGCN. The input layer prepares an adjusted transition matrix $\tilde{P}$ and a feature matrix of a given network, and feeds them to the first hidden layer. The hidden layers are composed of a multi-layer Graph Convolutional Network. The output layer has two neurons fullly connected by the last hidden layer, and each neuron corresponds to a class (0 or 1). To predict the probability distribution $Y^{'}$, i.e., the probability of a node belonging to each class, softmax is applied to the output layer.  The cross entropy of $Y^{'}$ and $Y$ is taken as the loss to train the model.}
  \label{1}
\end{figure}
\noindent The framework of FNGCN is shown in Fig.1. Similar to common GNNs, there are mainly three components in FNGCN, which are input layer, hidden layers and output layer. The main difference is that we construct centralities for each node by feature network, see the input layer component in Fig. 1. Next we briefly present the three components of FNGCN one by one.

\noindent \textbf{Input layer.} Given a network, the function of input layer is to prepare input data for FNGCN. There are two types of input data, one is the $ \tilde{P}=\tilde{D}^{-1/2}\tilde{A}\tilde{D}^{-1/2}$ matrix of the network, and the other one is the feature matrix. The $ \tilde{P}$ can be easily obtained from the adjcency matrix $\tilde{A} = A+I$ and the diagonal degree matrix $\tilde{D} = D + I$, i.e. The feature matrix contains the structural centralities of nodes, and each row corresponds to a node and each column corresponds to a centrality. To construct a proper feature matrix, we explore the structural centrality space in the view of network and apply the technique of network analysis, which will be described in details in Section 3.2. 

\noindent \textbf{Hidden layers.} Given the representation of the $\ell-th$  hidden layer $H^{\ell}$ , the $H^{\ell+1}$ commonly obtained as follows:
\begin{equation}\label{eqution1}
  H^{\ell+1} = \sigma(\tilde{P}H^\ell W^\ell)
\end{equation}
where  $W^\ell$ is the trainable weight matrix, $H^\ell$ = X when $\ell$ = 0, and $\sigma$ is the activation function. Later, Chen et al. \cite{2020GCNII} improve Eq. (1) and obtain $H^{\ell+1}$ as:
\begin{equation}\label{eqution2}
  H^{\ell+1}  = \sigma(((1-\alpha_\ell)\tilde{P}H^\ell+\alpha_\ell H^0)((1-\beta_\ell)I_n+\beta_\ell W^\ell))
\end{equation}
where $I_n$ is an identity matrix, $\alpha_\ell$ and $\beta_\ell$ are two hyperparameters. $\alpha_\ell$ ensures that the representation of $(\ell+1)-th$ layer retains at least part of the initial representation , and $\beta_\ell$  ensures that the decay of the weight matrix increases adaptively as the model stacks more layers. According to \cite{2020GCNII}, Eq. (2) can alleviate various problems such as over-smoothing caused by Eq. (1) when the hidden layers increase. In our work, to compare the performances of shallow GCNs and deep GCNs on the task of identifying influential nodes, we consider two settings for hidden layers. The first one only has 3 hidden layers and uses Eq. (1) to obtain the representations of hidden layers, which is denoted as FNGCN3; the second one has 64 layers and uses Eq. (2) to obtain the representations of hidden layers. Since Eq. (2) allows the representation of $(\ell+1)-th$ layer to retain at least part of the initial representation $H^0$, a fully connected layer is introduced between the input layer and the first hidden layer to transform the feature matrix into the same dimensions as the hidden layers, i.e., obtain $H^0$ \cite{2020GCNII}. It is noted that $H^0$ is actually the input feature matrix for Eq. (1). Moreover, to thoroughly explore the impact of shallow GCNs and deep GCNs on FNGCN, we will test FNGCN with 3, 8, 16, 24, 32 and 64 hidden GCN layers in the experimental part (details are referred to Section IV-D).

\noindent \textbf{Output layer.} A full-connected layer follows the last hidden layer for task learning. After that, a softmax layer is connected to obtain the probality of 
being influential for each node and normalize the probabilities, guaranteeing the probabilities summing to 1. To solve the problem of overflow and underflow of softmax, we apply Logsoftmax on the basis of softmax. Finally, we use the following cross entropy as the loss function: 
\begin{equation}\label{eqution8}
  L = -[Y\log_{}{\hat{Y}}+(1-Y)\log_{}{(1-\hat{Y})}]
  \end{equation}
  where $Y$ is the true label value and $\hat{Y}$ is the predicted probability value. It characterizes the difference between the true sample label and the predicted probability.
By optimizing the cross entropy loss on the train data, the parameter including $W^\ell$ can be computed, and the model is obtained. 
\subsection{Feature network} \label{subsec:centrality Construction}

\begin{table*}[ht]
  \centering
  \Huge
  \caption{The centralities considered as centralities and their definitions}
  \label{tab:definitions}
  \setlength{\tabcolsep}{3pt} 
	\renewcommand\arraystretch{1.5} 
  \resizebox{\columnwidth}{!}{%
  \begin{tabular}{c|c|c|l}
  \hline
   &
    Centrality &
    Definition &
    \multicolumn{1}{c}{Description} \\ \hline
  \multirow{4}{*}{Global centrality} &
    Closeness &
    $C_{c}\left(v_{i}\right)=\frac{n-1}{\sum_{j \neq i} d i s t_{i j}}$ &
    where $dist_{ij}$ is the shortest path between nodes $v_i$ and $v_j$ \\ \cline{2-4} 
   &
    Betweenness &
    $C_{b c}\left(v_{i}\right)=\sum_{s \neq i \neq t} \frac{\sigma_{s t}^{i}}{\sigma_{s t}}$ &
    \begin{tabular}[c]{@{}l@{}}where $ \sigma_st$ is the number of the shortest paths between nodes $v_s$ and $v_t$; \\ and $ \sigma_st^i$ is the number of the shortest paths that pass node $v_i$.\end{tabular} \\ \cline{2-4} 
   &
    PageRank &
    $C_{p r}\left(v_{i}\right)=\frac{1-\alpha}{n}+\alpha \sum_{v_{j} \in N\left(v_{i}\right)} \frac{C_{p r}\left(v_{j}\right)}{d_{j}}$ &
    \begin{tabular}[c]{@{}l@{}}where $\alpha$ is the scaling factor which is used to avoid that the nodes with \\ no neighbors absorb all the scores; $N(v_i )$ is the neighbors of node $v_i$;\\ $d_j$ is the degree of node $v_j$.\end{tabular} \\ \cline{2-4} 
   &
    Eigenvector &
    $C_{ec}(i)=c \sum_{j=1}^{n} a_{i j} x_{j}$ &
    \begin{tabular}[c]{@{}l@{}}where $c$ is a constant, $a_{ij}$ is a matrix element indicating that node $i$ is \\ connected to node $j$, and $x_j$ is the node $j$ eigenvector centrality value.\end{tabular} \\ \hline
  \multirow{11}{*}{Local centrality} &
    Degree &
    $d(u) = \frac{degree(u)}{n-1}$ &
    where $n$ is the number of nodes. \\ \cline{2-4} 
   &
    Extended Degree &
    $extd(u) = d_u+ {\textstyle \sum_{v\in N(u)}}^{} d_v$ &
    \begin{tabular}[c]{@{}l@{}}where $d_u$ is degree of node $u$; $N (u) $is the neighbor of node $u$;\\ $d_v$ is degree of node $v$.\end{tabular} \\ \cline{2-4} 
   &
    Accumulated Degree &
    $accd(u) = d_u+ {\textstyle \sum_{v\in N(u)}}^{}(d_v+ {\textstyle \sum_{w\in N(v)}^{}}d_w )$ &
    \begin{tabular}[c]{@{}l@{}}where $d_u$ is degree of node $u$, $N (u) $is the neighbor of node $u$;\\ $d_v$ is degree of node $v$,$d_w$ is degree of node $w$.\end{tabular} \\ \cline{2-4} 
   &
    Node Mass &
    $nm(u)=\mid {(v,w)\in E\mid v,w\in N_u^+}\mid$ &
    \begin{tabular}[c]{@{}l@{}}where $E$ and $V$ are the set of edges and the set of nodes\\ of network,respectively; $N(u)^+$ is set of the neighbor nodes \\ of $u$, including node $u$.\end{tabular} \\ \cline{2-4} 
   &
    Conductance of Egonet &
    $\operatorname{conductance}(u)=\frac{\left|\left\{(v, w) \in E \mid v \in N^{+}(u), w \in V-N^{+}(u)\right\}\right|}{\min \left(\operatorname{vol}\left(N^{+}(u)\right), \operatorname{vol}\left(V-N^{+}(u)\right)\right)}$ &
    \begin{tabular}[c]{@{}l@{}}where $E$ and $V$ are the set of edges and the set of nodes\\ of network,respectively; $N(u)^+$ is set of the neighbor nodes \\ of $u$, including node $u$; $V-N(u)^+$ is the nodes ecpect node $u$\\ and its neighbors; $val()$ is number of nodes in the set.\end{tabular} \\ \cline{2-4} 
   &
    Density of Egonet &
    $\operatorname{density}(u)=\frac{\left|\left\{(v, w) \in E \mid v, w \in N^{+}(u)\right\}\right|}{\left(\begin{array}{c} d_{u}+1 \\ 2\end{array}\right)}$ &
    \begin{tabular}[c]{@{}l@{}}where $E$ is the set of edges of network,respectively;\\ $N(u)^+$ is set of the neighbor nodes of $u$, including node $u$; \\ $d_u$ is the degree of node $u$.\end{tabular} \\ \cline{2-4} 
   &
    \begin{tabular}[c]{@{}c@{}}Local Clustering \\ Coefficient(LCC)\end{tabular} &
    $LCC(u)=\frac{|\{(v,w)\in E|v,w\in N(u)\}|}{\left(\begin{smallmatrix}d_{u}\\2\end{smallmatrix}\right)}$ &
    \begin{tabular}[c]{@{}l@{}}where $E$ is the set of edges of network;\\ $N(u)$ is set of the neighbor nodes of $u$; \\ $d_u$ is the degree of node $u$.\end{tabular} \\ \cline{2-4} 
   &
    \begin{tabular}[c]{@{}c@{}}Core Dominance with\\  Cosine index(CoredCosine)\end{tabular} &
    $\operatorname{sim}_{\text {Cosine }}(u, v)=\frac{|N(u) \cap N(v)|}{\sqrt{|N(u)||N(v)|}}$ &
    \begin{tabular}[c]{@{}l@{}}where $N(u)$ is set of the neighbor nodes of $u$ and $N(v)$ \\ is set of the neighbor nodes of $v$.\end{tabular} \\ \cline{2-4} 
   &
    \begin{tabular}[c]{@{}c@{}}Core Dominance with \\  Jaccard index(CoedJaccard)\end{tabular} &
    $\operatorname{sim}_{\text {Jaccard }}(u, v)=\frac{|N(u) \cap N(v)|}{|N(u)\cup N(v)|}$ &
    \begin{tabular}[c]{@{}l@{}}where $N(u)$ is set of the neighbor nodes of $u$ and $N(v)$ \\ is set of the neighbor nodes of $v$.\end{tabular} \\ \cline{2-4} 
   &
    \begin{tabular}[c]{@{}c@{}}Core Dominance with \\  Pearson Correlation \\  Coefficient(CoredPearson)\end{tabular} &
    $\operatorname{sim}_{\text {Pearson }}(u, v)=\frac{\sum_{k}\left(A_{u, k}-\bar{A}_{u}\right)\left(A_{v, k}-\bar{A}_{v}\right)}{\sqrt{\sum_{k}\left(A_{u, k}-\bar{A}_{u}\right)^{2}} \sqrt{\sum_{k}\left(A_{v, k}-\bar{A}_{v}\right)^{2}}}$ &
    where $A$ is the adjacency matrix, and $\bar{A}$ is the average degree. \\ \cline{2-4} 
   &
    \begin{tabular}[c]{@{}c@{}}Core Dominance with \\  Preference Attachment(SPA)\end{tabular} &
    $sim_{PA}(u,v) = d_u \times d_v$ &
    where $d_u$ and $d_v$ is the degree of node $u$ and the degree node of $v$,respectively. \\ \hline
  \end{tabular}%
  }
  \end{table*}
\noindent In network analysis, "centrality" is a very important concept developed to identify important nodes in a network. Therefore, it is widely used to identify influential nodes in various networks in the literature \cite{2022Comprehensivecentality,2019Rankingnodes}. However, "importance" has a wide number of meanings, leading to many different definitions of centrality (centrality-based methods.), see TABLE I for examples. Those centrality metrics assess the importance of nodes from different perspectives. For example, the eigenvector and PageRank centralities counts the number of walks of length infinity starting and ending from a node, the closeness centrality considers the shortest distance from a node to all other nodes and the betweenness centrality counts the shortest pathes passing through a node. As the increase in network data, many local centralities are also proposed to assess the importance of nodes by only considering the local structure of nodes, of course, from different angles. For example, the degree counts the neighbors of a node, while the Extended Degree \cite{2010exd} and Accumulated Degree \cite{2005acd} consider the degrees of a node's neighbors on the basis of its only degree; the Node Mass \cite{2018nd}, Condctance of Egonet \cite{2016ce}, Density of Egonet  and LCC \cite{2015lcc} centralities evaluate the closeness of a node's neighborhood; other local centralities, such as the CoredCosine, CoredJaccard, CoredPeasron and CoredSPA  \cite{2018dom} centralities, consider the similarity between a node and its neighbors.  It is therefore incomplete to apply one or two centralities to identify influential nodes. A recently popular approach is to take several centralities as node centralities, and input the node centralities with the network to a graph neural network model to identify influential nodes \cite{2020inf} \cite{2023basedcentralitymeasuresandcentralityselection}. However, there are many centralities defined from different perspectives, and some of them have very similar meanings, as shown in TABLE I. In addition, it is computationally intensive to calculate the global centralities, especially for large networks. Thus, it is important to choose the most related and irredundant centralities to construct the node features, and explore whether the local centralities are enough to identify influential nodes.  

\noindent Based on the discussions above, we propose to construct the node features from the view of network analysis. It has been demonstrated that network is an intuitive representation of complex systems, and many techniques of network analysis are available. The hidden patterns in a complex system can be more easier to uncover from the view of network analysis. In particular, we construct the node features as Algorithm 1. Given a network, the first step is to calculate the local centrality values of each node in the network (see step S1). Here we only consider local centralities because of efficiency, and in the experimental part we will conduct an experiment to demonstrate that local centralities are enough to identify influential nodes. A feature network is then constructed according to the local centrality values (see step S2). Specifically, each centrality is represented as a node named centrality node, and the degree of relevance between each pair of centralities is evaluated by the Spearman  Correlation  Coefficient \cite{2004spearman}:
{\small
\begin{equation}
  \operatorname{Scc}(c 1, c 2)=\frac{\frac{1}{n} \sum_{i=1}^{n}\left(x_{i}-\bar{x}\right) \cdot\left(y_{i}-\bar{y}\right)}{\sqrt{\left(\frac{1}{n} \sum_{i=1}^{n}\left(x_{i}-\bar{x}\right)^{2}\right) \cdot\left(\frac{1}{n} \sum_{i=1}^{n}\left(y_{i}-\bar{y}\right)^{2}\right)}}
\end{equation}
}
where $c1$ and $c2$ are centrality nodes corresponding to two centrality metrics; $n$ is the number of nodes in the network, $x_{i}$ is the ordinal number of the $i-th$ node after sorting according to the centrality value calculated by $c1$, and $y_{i}$ is the ordinal number of the $i-th$ node after sorting according to the centrality value calculated by $c2$. It is noted that here we use Spearman Correlation Coefficient, rather than other similarities such as Cosine similarity, Jaccard similarity and pHash simialrity\cite{2024sd}, because it is insensitive to variables with different scales.  If $Scc(c1, c2)$ is larger than $\delta$ (a given threshold), then there is an edge between $c1$ and $c2$. According to the structure of the feature network, the centralities are clustered into different groups (see step S3), which can be fulfilled by any community detection algorithm.  Because the centralities in a group are highly correlated, we only choose one centrality from each group. To keep as much information as possible and minimize redundancy between the information, we choose the centrality node with the largest degree in its belonging group and no neighbors are chosen yet (see step S4). Finally, all chosen centralities are taken as centralities, and the node features is composed of the values of centralities on each node (corresponding to a row in the node features) (see step S5). To avoid overfitting, we normalize the values of each centrality \cite{2020inf} as following:
\begin{equation}
f_k(c) = \frac{R_k(c)}{N}-0.5 
\end{equation}
where $R_k(c)$ is the rank of node $k$ in descending order of the centrality $c$ (i.e., the chosen centrality c), and $N$ is the number of nodes.
\begin{table}[!htb]
\centering
\caption{Algorithm of  construction of node features}
\label{tab:Algorithm 1}
\resizebox{\columnwidth}{!}{%
\begin{tabular}{l}
\hline
Algorithm 1 the construction of node features                                                       \\ \hline
Input: G=(V,E): a network; $\delta$: the threshold to connect two centrality nodes.                     \\ 
Output: the node features                                                                           \\ 
S1. Calculate the local centrality values of each node in the network.                               \\ 
\begin{tabular}[c]{@{}l@{}}S2. Construct feature network:\\ 1) take each centrality as a node named centrality node;\\ 2)for each pair of centrality node $c1$ and $c2$: if $Scc(c1,c2)$ \textgreater{} $\delta$, then add an adge to $c1$ and $c2$.\end{tabular} \\ 
S3. Cluster centrality nodes into different groups according to the structure of feature network. \\ 
\begin{tabular}[c]{@{}l@{}}S4. Choose from each group a centrality satisfying: the centrality node has the largest degree in its\\ belonging group and there is no neighbor centrality node chosen yet.\end{tabular} \\ 
\begin{tabular}[c]{@{}l@{}}S5. Apply the chosen centralities as the features for each node, and normalize each feature to obtain\\ the node features.\end{tabular} \\ \hline
\end{tabular}%
}
\end{table}
\noindent  For illustration, we take CA-CondMat as an example. After obtaining the local centrality values of each node in CA-CondMat, we calculate the Scc between each pair of local centralities and construct the feature network (here $\delta$  is set to 0.9), see Fig.2. According to the feature network, these local centralities can be clustered into five groups, and they are {Accumulated Degree, Extended Degree, SPA}, {Degree, Node Mass, CoredPearson, CoredCosine, CoredJaccard}, {Density of Egonet}, {Conductance}, {LCC}. Because the SPA has the largest degree which is 4, the SPA is chosen to represent the first group. In the second group, the Node Mass has the largest degree which is 5, but its neighbor the SPA has been chosen, thus it is not chosen. CoredCosine and CoredPearson have the second largest degree which is 4, and there is no neighbor of them to be chosen. In this case, we can choose either of them. Each of the remaining three groups has only one centrality, and all of them are chosen.  Therefore, the features for CA-CondMat are SPA, CoredCosine, Density of Egonet, Conductance and LCC, assuming CoredCosine is chosen to represent the second group. After the features are determined, the feature matrix is obtained by taking the normalized feature values of each node, i.e., the normalized values of each node on the chosen centralities. The construction of feature network is the same for all datasets, and the chosen centralities for the six networks are shown in TABLE III.
\begin{figure}[ht]
  \centering
  \includegraphics[height=4cm,width=7.8cm]{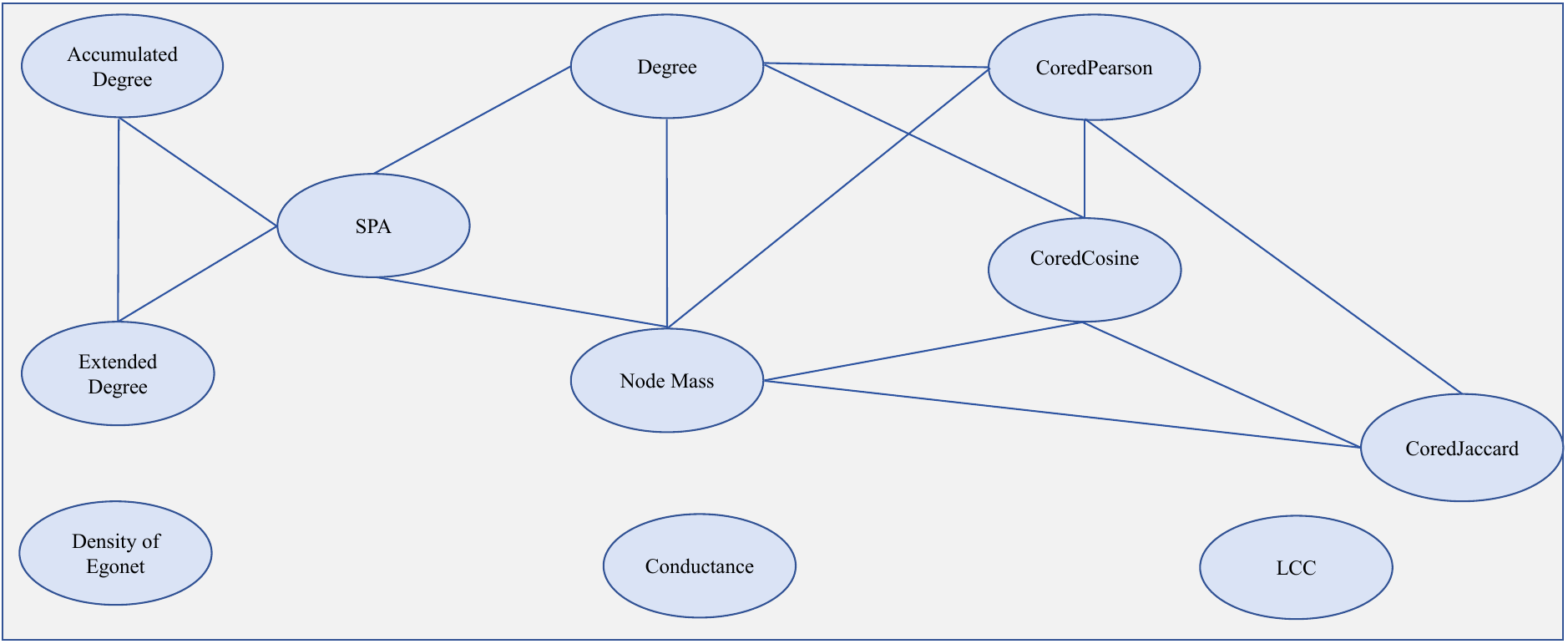}
  \caption{The feature network of CA-CondMat}
  \label{fig2}
\end{figure}
\par
\begin{table}[ht]
  \centering
  \caption{Local centralities for each network}
  \normalsize
  \label{tab:Local centralities for each network}
  \resizebox{\columnwidth}{!}{%
  \begin{tabular}{c|l}
  \hline
  Network             & \multicolumn{1}{c}{Centrality}                                                                                   \\ \hline
  Hamsterster Friends,Sister-Cities, CA-CondMat & \begin{tabular}[c]{@{}l@{}}SPA, CoredCosine,\\ CoredPearson, Density of \\ Egonet,Conductance, LCC\end{tabular}  \\ \hline
  Human protein (Vidal) &
    \begin{tabular}[c]{@{}l@{}}SPA, CoredCosine,\\ CoredPearson, Density of\\ Egonet, Conductance,\\ LCC, Degree\end{tabular} \\ \hline
  \begin{tabular}[c]{@{}c@{}}CA-GrQc, CA-HepTh\end{tabular} &
    \begin{tabular}[c]{@{}l@{}}SPA, CoredCosine,\\ Density of Egonet, \\ Conductance, LCC\end{tabular} \\ \hline
         
  \end{tabular}%
  }
  \end{table}

\subsection{SIR model} 
\label{subsec:SIR model}
\noindent Similar to other GNNs used to identify influential nodes, FNGCN is a supervised model which needs labeled data to train. However, it is usually unknown which nodes are influential in a given network. To label the data, we apply the SIR model to score the influence of each node in a network, rank nodes in descending order of the influence score, and take the top $5\%$ nodes as influential nodes.

\noindent The SIR model is originally proposed to predict the transmissibility of infectious diseases, and later, it is widely used for the dissemination of information, and a node's influence in a network can be quantified by the spreading area of the information starting from that node \cite{2020covidSIR,2021SIRinfluence}. In the SIR model, there are three types of nodes: the susceptible nodes, the infected nodes, and the recovered nodes. A susceptible node can be infected by its infected neighbors with a probability named infection rate $\beta$, and become an infected node. An infected node can transfer to a recovered node with a probability named recovery rate $\gamma$. Once a node becomes a recovered node, it cannot be infected again. When using the SIR model to quantify the influence of a node, we initially set this node as an infected node and the other nodes as susceptible nodes. Then the infected node infects its neighbors, and its neighbors in turn infect their neighbors, of course, with $\beta$. Moreover, the infected nodes can transfer to recovered nodes with $\gamma$. This process continues until there is no more nodes to become infected, and the number of infected nodes and recovered nodes is taken to score the influence of the initial infected node as follows:
\begin{equation}\label{eq4}
  IC = \frac{{\sum_{j=1}^{1000}(R^j+I^j)}}{N\times1000}
\end{equation}
\newline
where $IC$ is the influence capability of a node, $N$ is the total number of nodes in the network, and $R^j$ and $I^j$ are the numbers of recovered nodes and infected nodes in the $j-th$ experiment, respectively. Because an infected node both infects its susceptible neighbors and transfer to a recovered node with a probability, the process of infection is a stochastic process. To reduce stochasticity, for each node we independently perform the SIR model 1000 times, and take the average value of all ICs as the final influence score. 

\noindent In addition, we follow \cite{2022cgcn} to set the recovery rate $\gamma$ and the infection rate $\beta$. In particular, $\gamma$ is set to 1, and $\beta$ = $\xi$ $\times$ $\beta_c$ where $\xi=2$ and $\beta_c$ is obtained according to virtue of the mean-field theory and is calculated based on the average degree $<d>$ of the network, namely:
\begin{equation}\label{eq5}
  \beta_c = \frac{<d>}{{< d^2>}- <d>}
\end{equation}
\section{Experiment} \label{sec:Experiment}
\noindent In this section, we conduct three experiments to test the proposed FNGCN. The first one is to compare FNGCN with other methods. Then, we conduct four ablation experiments aiming at exploring the following key issues: the impact of the number of hidden GCN layers on FNGCN, verifying that local features are enough for GCN to identify influential nodes, it is necessary to select proper local features, and the contributions of different local centralities to influential node identification. Next, we first describe the datasets used in the experiments, then present the experimental setting and evaluation indicators, and finally present the experimental results and anlysis.
\subsection{Datasets} \label{subsec:Datasets}
\noindent We use six real-world networks in the experiments, see TABLE IV for the statistical information of these networks. Hamsterster Friends contains friendships between users of the Hamsterster website; Human protein (Vidal) reproduces an initial version of the proteome map of human binary protein-protein interactions; CA-GrQc is a collaboration network in the field of "general relativity"; CA-HepTh is a collaboration network in the field of "high energy physics theory"; and CA-CondMat is a collaboration network in the field of "condensed matter", all obtained from Arxiv; Sister-Cities is an undirected network representing the "sister cities" or "twin cities" relationships between cities.
\begin{table}[ht]
  \centering
  \caption{The statistical information of the used networks}
  \label{tab:The statistical information of the used networks}
  \resizebox{\columnwidth}{!}{%
  \begin{tabular}{c|c|c|c|c|c|c}
  \hline
  Network & Nodes & Edges & Diameter & Average Degree & Average Clustering Coefficient & \multicolumn{1}{c}{Source} \\ \hline
  Hamsterster Friends   & 1858  & 12534 & 14 & 13.96 & 0.14  & \cite{2017d} \\ \hline
  Human protein (Vidal) & 3133  & 6726  & 13 & 4.32  & 0.072 & \cite{2017d} \\ \hline
  CA-GrQc               & 4158  & 13422 & 17 & 6.46  & 0.56  & \cite{2007d} \\ \hline
  CA-HepTh              & 9877  & 25998 & 18 & 5.74  & 0.48  & \cite{2007d} \\ \hline
  Sister-Cities         & 14274 & 20573 & 25 & 3.49  & 0.048 & \cite{2017d} \\ \hline
  CA-CondMat            & 23133 & 93497 & 15 & 8.55  & 0.64  & \cite{2017d} \\ \hline
  \end{tabular}%
  }
\end{table}
\subsection{Experimental setting and Evaluation} \label{subsec:Experimental setting and Evaluation}
\noindent We consider two architectures for FNGCN: one is a shallow graph convolution network containing only 3 hidden layers and 1 fully connected layers, named FNGCN3, and the other one is a deep graph convolution network containing 64 hidden layers and two fully connected layers, named FNGCN64. For FNGCN3, the weight decay and dropout rate are set to 0.01 and 0.4; for FNGCN64, the weight decays for hidden layers and fully connected layers are set to 0.01 and 0.0005, respectively, and the dropout is set to 0.4. The learning rate for FNGCN3 and FNGCN64 are set to 0.01 and 0.05, respectively, and the Adam optimizer\cite{2014ad} is taken as optimization algorithm. The hyperparameters $\alpha_l$ and $\lambda$ in $\beta_\ell$ are set to 0.1 and 0.5. Recall the labeling of data in Section 3.3, only the top $5\%$ nodes in a network are labeled the influential nodes, while the remaining $95\%$ nodes are not. To address the issue of serious imbalance, for each network we sample $10\%$ non-influential nodes proportionally to the influence score, and construct a more balanced dataset with the ratio between positive and negative to be 1:2. Then $70\%$ nodes of the obtained dataset are sampled as training set and the remaining nodes form the test set. In the training phase, we apply early stopping to avoid overfitting. In particular, the training will be terminated  when the loss on the test set is continuing to deteriorate for 30 and 100 epochs on FNGCN3 and FNGCN64 respectively, and the model is obtained by taking the one corresponding to the best loss, i.e., the one obtained at the first epoch that loss of the test set is not decreasing. To avoid
over-fitting, we also use DropOut technology in each hidden layer and early stopping technology.

\noindent To evaluate the results of each model, we apply a variety of three evaluation indicators including accuracy, F1-score and Area Under Curve (AUC). Accuracy is the ratio of the samples correctly classified by the classifier to the total samples, F1-score \cite{2014f1} is the harmonic mean of precision and recall. Precision is the proportion of samples with positive predicted outcomes that are actually positive and recall is the proportion of samples with positive predicted outcomes that are actually positive to the proportion of positive samples in the full sample. The AUC \cite{1997auc} is the area enclosed with the coordinate axis under the ROC curve, which is a curve plotted with the true positive rate as the vertical coordinate and the false positive rate as the horizontal coordinate.
\begin{table}[!htb]
  \centering
  \caption{Results of different methods on different datasets}
  \resizebox{\columnwidth}{!}{%
  \begin{tabular}{c|c|c|c|c|c|c|c}
  \hline
  Method &
    \multicolumn{1}{l|}{} &
    \begin{tabular}[c]{@{}c@{}}Hamsterster\\ friends\end{tabular} &
    \begin{tabular}[c]{@{}c@{}}Human \\ Proteins\\  (Vidal)\end{tabular} &
    \begin{tabular}[c]{@{}c@{}}CA-\\ GrQc\end{tabular} &
    \begin{tabular}[c]{@{}c@{}}CA-\\ HepTh\end{tabular} &
    \begin{tabular}[c]{@{}c@{}}Sister-\\ Cities\end{tabular} &
    \begin{tabular}[c]{@{}c@{}}CA-\\ CondMat\end{tabular} \\ \hline
  SVM                          & \multirow{7}{*}{Acc} & 0.926 & 0.944 & 0.926 & \textbf{0.964} & 0.970 & 0.972 \\ \cline{1-1} \cline{3-8} 
  LR                           &                      & 0.926 & 0.937 & 0.915 & 0.951 & 0.961 & 0.956 \\ \cline{1-1} \cline{3-8} 
  InfGCN                       &                      & -     & -     & -     & -     & -     & -     \\ \cline{1-1} \cline{3-8} 
  GATv2                        &                      & 0.951 & 0.929 & 0.931 & 0.913 & 0.940 & 0.932 \\ \cline{1-1} \cline{3-8} 
  GATv2-FN                     &                      & 0.963 & 0.944 & 0.926 & 0.908 & 0.940 & 0.947 \\ \cline{1-1} \cline{3-8} 
  FNGCN64                      &                      & \textbf{0.975} & 0.952 & \textbf{0.952} & \textbf{0.964} & \textbf{0.981} & 0.967 \\ \cline{1-1} \cline{3-8} 
  FNGCN3                       &                      & \textbf{0.975} & \textbf{0.968} & 0.936 & 0.956 & 0.979 & \textbf{0.978} \\ \hline
  SVM                          & \multirow{7}{*}{F1}  & 0.900 & 0.923 & 0.899 & \textbf{0.949} & 0.957 & 0.959 \\ \cline{1-1} \cline{3-8} 
  LR                           &                      & 0.900 & 0.911 & 0.886 & 0.932 & 0.945 & 0.938 \\ \cline{1-1} \cline{3-8} 
  InfGCN                       &                      & 0.903 & 0.700 & 0.859 & 0.907 & -     & 0.814 \\ \cline{1-1} \cline{3-8} 
  GATv2                      &                      & 0.931 & 0.894 & 0.893 & 0.866 & 0.915 & 0.900 \\ \cline{1-1} \cline{3-8} 
  GATv2-FN                     &                      & 0.941 & 0.920 & 0.887 & 0.859 & 0.914 & 0.922 \\ \cline{1-1} \cline{3-8} 
  FNGCN64                      &                      & \textbf{0.964} & 0.932 & \textbf{0.931} & 0.948 & \textbf{0.972} & 0.952 \\ \cline{1-1} \cline{3-8} 
  \multicolumn{1}{l|}{FNGCN3} &                      & \textbf{0.964} & \textbf{0.955} & 0.908 & 0.938 & 0.969 & \textbf{0.968} \\ \hline
  SVM                          & \multirow{7}{*}{AUC} & 0.998 & 0.991 & \textbf{0.981} & \textbf{0.997} & 0.993 & 0.992 \\ \cline{1-1} \cline{3-8} 
  LR                           &                      & 0.994 & 0.990 & 0.978 & 0.992 & 0.995 & 0.993 \\ \cline{1-1} \cline{3-8} 
  InfGCN                       &                      & 0.981 & 0.939 & 0.974 & 0.981 & -     & 0.955 \\ \cline{1-1} \cline{3-8} 
  GATv2                        &                      & 0.977 & 0.958 & 0.979 & 0.971 & 0.973 & 0.975 \\ \cline{1-1} \cline{3-8} 
  GATv2-FN                     &                      & 0.989 & 0.976 & 0.976 & 0.972 & 0.978 & 0.982 \\ \cline{1-1} \cline{3-8} 
  FNGCN64                      &                      & \textbf{0.999} & \textbf{0.993} & 0.972 & 0.996 & 0.996 & 0.995 \\ \cline{1-1} \cline{3-8} 
  \multicolumn{1}{l|}{FNGCN3} &                      & \textbf{0.999} & \textbf{0.993} & 0.969 & 0.996 & \textbf{0.997} & \textbf{0.997} \\ \hline
  \end{tabular}%
  }
  \tiny {1. The accuracy and the results on Sister-Cities by InfGCN is not provided by the original paper, and is marked as “-”}
  \end{table}

\subsection{Comparison between FNGCN and other methods} \label{subsec:Comparison between FNGCN and other methods}
\noindent In this experiment, we compare FNGCN3 and FNGCN64 with two traditional classification models Support Vector Machines (SVM), Logistic Regression model (LR),
and two GNN-based models which are InfGCN and GATv2\cite{2022GATv2}. InfGCN is an excellent GNN-based model for influential node identification. For GATv2, we consider two versions: one version takes common centralities same as InfGCN as node features (denoted as GATv2 in the following text), and the other version selects local centralities as described in Section III-B as node features (denoted as GATv2-FN in the following text). For consistency, we use the results of InfGCN provided by the original paper\cite{2020inf}, and use the implementation of GATv2 provided by the authors \cite{2022GATv2}. The results of different models are shown in Table V.

\noindent \textbf{The comparison between FNGCN and other methods.} According to table 5, FNGCNs outperforms SVM in all cases, except that SVM performs better on CA-HepTh, and leads to better AUC than FNGCNs on CA-GrQc, better accuracy and F1-score than FNGCN64 on CA-CondMat. Compared with InfGCN, LR, GATv2, and GATv2-FN, FNGCNs performs better across all the datasets except CA-GrQc, where FNGCNs lead to a worse AUC. Further, compared with the improvement, the degradation by FNGCNs on CA-GrQc, CA-HepTh, and CA-CondMat is slighter. Specifically, compared with SVM, FNGCNs lead to an improvement of 5.3\%, 7.1\%, and 0.1\% in terms of accuracy, F1-score, and AUC, respectively, on the Hamsterster Friends; FNGCN3 leads to an improvement of 2.5\%, 3.5\%, and 0.2\% in terms of accuracy, F1-score, and AUC, respectively, on the Human Proteins (Vidal), an improvement of 0.4\% in terms of AUC on Sister-Cities, an improvement of 0.6\%, 0.9\%, and 0.5\% in terms of accuracy, F1-score, and AUC, respectively, on CA-CondMat; FNGCN64 leads an improvement of 2.8\% and 3.6\% in terms of accuracy and F1-score, respectively, on CA-GrQc, an improvement of 1.1\% and 1.4\% in terms of accuracy and F1-score, respectively, on Sister-Cities. Compared with SVM, FNGCN3 leads to a degradation of 0.8\%, 1.2\%, and 0.1\% in terms of accuracy, F1-score, and AUC respectively, on CA-HepTh, a degradation of 0.5\% and 0.7\% in terms of accuracy and F1-score, respectively, on CA-CondMat; FNGCN64 leads to a degradation of 0.1\% in terms of both F1-score and AUC on CA-HepTh. On CA-GrQc, compared with LR, InfGCN, GATv2, and GATv2-FN, FNGCN64 (FNGCN3) leads to degradations of 0.6\% (0.9\%), 0.2\% (0.5\%), 0.7\% (1.0\%), and 0.4\% (0.7\%), respectively, in terms of AUC. Moreover, GATv2-FN outperforms GATv2 in most cases. These results demonstrate the effectiveness of the proposed FNGCN on identifying influential nodes.
\subsection{Ablation Experiments} \label{subsec:Ablation Experiments}
\noindent  To further verify the FNGCN, we conduct four ablation experiments to test the effect of the approach to constructing node features in FNGCN. The first one is  to explore the impact of the number of hidden GCN layers on FNGCN; the second one is to test whether it is feasible to only use local centralities as features, since local centralities are much easier to calculate; the third one is to test the centrality selection approach in FNGCN;  the last one is the contributions of different local centralities to influential node identification. Therefore, in the first ablation experiment, we constructs 3-layer, 8-layer, 16-layer, 24-layer, 32-layer and 64-layer FNGCN, respectively; in the second ablation experiment, we compare GCN with  all centralities, global centralities and local centralities, i.e., GCN with all local centralities to construct node features, GCN with all global centralities to construct node features and the ones with all centralities, including the local centralities and global centralities, to construct node features (see Table for the centralities); in the third ablation experiment, we compare FNGCN with  GCN that use all local centralities to construct node features and GCN with inputs of InfGCN to construct node features, which includes degree, betweenness centrality, closeness centrality and clustering coefficien. For consistency, all the compared GCN are the same to FNGCN, except that the formers do not apply the approach presented in Section 3.2 to construct node features. In the fourth ablation experiment, we tests the contribution of different local centrality to the model based on the FNGCN3 and FNGCN64, respectively. Specifically, one type of local feature is deleted respectively, and then the features of the nodes are reconstructed based on the remaining local features only. With this approach, we evaluates the importance of these local features in identifying influential nodes and their ability to contribute to the model.

\begin{figure}[!htb]
  \centering 
  \includegraphics[height=1.8cm,width=8cm]{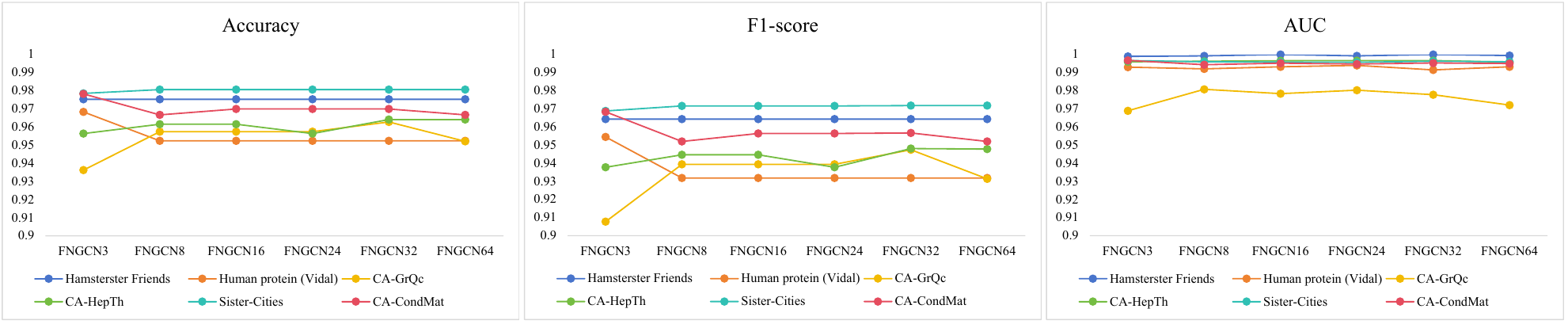}
  \caption{The results of  the impact of the number of hidden GCN layers on FNGCN}
  \label{3}
\end{figure}

\noindent \textbf{The impact of the number of hidden GCN layers on FNGCN.}
In order to further explore the impact of the number of hidden GCN layers (i.e., the hidden layers) on the identification model, we test FNGCN with 3, 8, 16, 24, 32 and 64 hidden GCN layers, respectively, and the results are shown in Fig. 3 (FNGCN with x layers is denoted as FNGCNx). According to Fig.3, the performance of FNGCN changes with different hidden GCN layers, but the change not significant. Particularly, on Hamsterster Friends, FNGCN with different hidden layers performs comparably in terms of all evaluation indicators. On Human protein (Vidal), the ranges of accuracy, F1-score, and AUC are [0.952,0.968], [0.931,0.955], and [0.992,0.994], respectively; in terms of accuracy and F1-score, FNGCN3 performs best, and the other FNGCNs perform almost equally; in terms of AUC, the result is similar, but a slight degradation is produced by FNGCN32. On CA-CondMa, the ranges of accuracy, F1-score, and AUC are [0.967,0.978], [0.952,0.968], and [0.994,0.997], respectively; in terms of accuracy and F1-score, FNGCN3 performs best, and then two slight degradations are produced by FNGCN8 and FNGCN64, respectively; in terms of AUC, FNGCN3 performs best, and the others performs almost equally. On CA-GrQc and CA-HepTh, the ranges of accuracy, F1-score, and AUC are respectively [0.936,0.962], [0.908,0.947], and [0.969,0.981] on CA-GrQc, and respectively [0.956,0.964], [0.938,0.948], and [0.996,0.995] on CA-HepTh; in terms of accuracy and F1-score, FNGCN3 performs worst, and FNGCN32 produces a slight improvement; moreover, on CA-GrQc, FNGCN64 produces a degradation, and on CA-HepTh, FNGCN24 produces a slight degradation. In terms of AUC, on CA-GrQc, FNGCN3 performs worst, and FNGCN8 produces an improvement and then decreased performance is observed overall along with the number of hidden layers increasing; on CA-HepTh, FNGCN with different hidden layers performs comparably. On Sister-Cities, the ranges of accuracy, F1-score and AUC are [0.979,0.981], [0.969,0.972], and [0.996,0.997], respectively; in terms of accuracy and F1-score, FNGCN with 8, 16, 24, 32 and 64 layers performs comparably, and is slightly better than FNGCN3; in terms of AUC, all FNGCNs performs comparably.
\par
From the results above, it can be inferred that: on Hamsterster Friends, the performance of FNGCN is almost not effected by the number of hidden GCN layers; on Human protrein (Vidal) and CA-CondMat, three hidden GCN layers produces the best performance, and the increasing hidden layers do not improve the performance, and even cause a degradation in some cases; on CA-GrQc, CA-HepTh and Sister-Cities, the increasing hidden layers bring improvement overall, but the improvement almost disappears (a degradation is probably observed in some cases) when the number of hidden layers is too large (over 8 here). Overall, deep GCNs indeed perform better than shallow GCNs in some cases, which is reasonable since deep GCN is able to aggregate information from more nodes; but shallow GCNs also perform well, and the difference between deep GCNs and shallow GCNs is not large, indicating that local information has the ability to distinguish nodes in most cases. These results imply that the impact of the number of hidden GCN layers on FNGCN is not significant.

\begin{figure}[!t]
  \centering
  \subfloat[]{\includegraphics[width=3in]{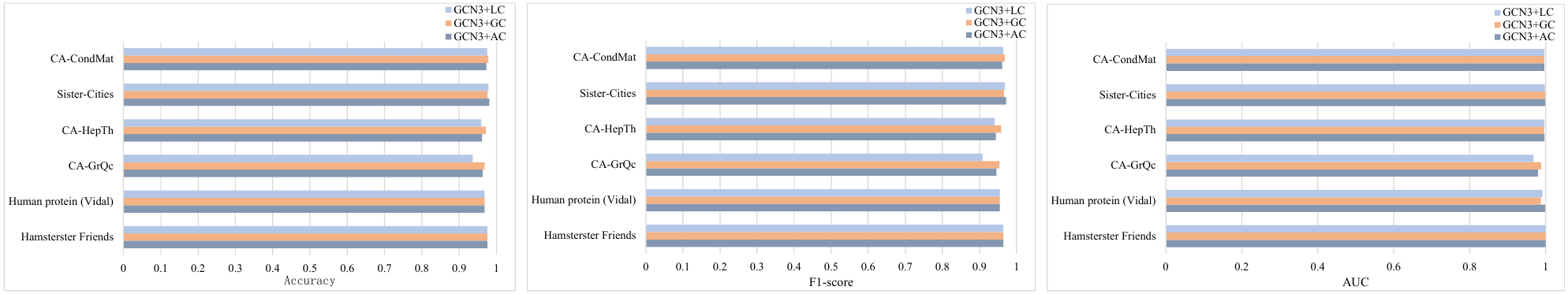}%
  \label{fig_first_case2}}
  \hfil
  \subfloat[]{\includegraphics[width=3in]{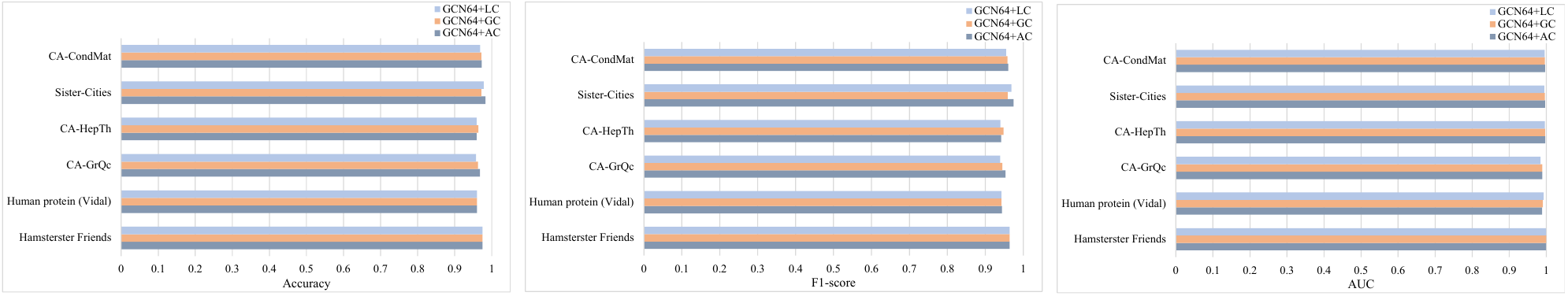}%
  \label{fig_second_case2}}
  \caption{The results of the first ablation experiments, i.e., comparing GCN with local centralities, GCN with global centralities and GCN with all centralities (including local and global centralities. (a) The results obtained by GCN3 with local centralities (abbreviated as GCN3+LC), GCN3 with global centralities (abbreviated as GCN3+GC) and GCN3 with all centralities (abbreviated as GCN3+AC, (b) The results obtained by GCN64 with local centralities (abbreviated as GCN64+LC),  GCN64 with global centralities (abbreviated as GCN64+GC) and GCN64 with all centralities (abbreviated as GCN64+AC)}
  \label{fig_sim2}
  \end{figure}

\noindent \textbf{Comparisons between local centralities,  global centralities, and  all centralities (including local and global centralities).} The results of the first ablation experiment are shown in Fig.4, and the results obtained by GCN with 3 hidden layers, named GCN3, are shown in Fig.4 (a) and the results on GCN with 64 hidden layers, named GCN64, are shown in Fig.4 (b). According to Fig.4 (a), compared with GCN3 with global centralities, GCN3 with local centralities leads to a little degradation on CA-CondMat, CA-HepTh and CA-GrQc, but performs equally on Human Protein (Vidal) and Hamsterster Friends, and even brings  improvement on Sister-Cities. Compared with GCN3 with all centralities, GCN3 with local centralities leads to a little degradation on Sister-Cities, CA-HepTh and CA-GrQc, but performs equally on Human Protein (Vidal) and Hamsterster Friends, and even brings slight improvement on CA-CondMat. According to Fig.4 (b), compared with GCN64 with global centralities, GCN64 with local centralities leads to slight degradation  on CA-CondMat, CA-HepTh and CA-GrQc in terms of  accuracy and F1-score, on CA-HepTh in terms of AUC,  but performs equally on  Human protein (Vidal), Hamsterster Friends, and CA-CondMat in terms of AUC,  and  even brings  improvement on Sister-Cities in terms of accuracy and F1-score, on  Sister-Cities and CA-GrQc in terms of AUC.Compared with GCN64 with all centralities, GCN64 with local centralities leads to slight degradation  on  CA-CondMat, Sister-Cities and CA-GrQc in terms of accuracy, on CA-CondMat, Sister-Cities, CA-GrQc, Human protein (Vidal) and CA-HepTh in terms of F1-score and AUC, but performs equally on  Hamsterster Friends, CA-HepTh and Human protein (Vidal)  in terms of accuracy. 
Moreover, according to TABLE VI, the calculation of local centralities costs much less time than  those of global centralities and all centralities on all networks. These results indicate that local centralities can lead to results comparable with global centralities and all centralities, but with much less time consumption, and even bring improvements in some cases, demonstrating that it is feasible to only use local centralities as node features for GCN. 
\begin{table}[htb]
  \centering
  \caption{Time consumption by the calculation of all centralities, global centrality and local centrality (in seconds)}
  \label{Time consumption}
  \resizebox{\columnwidth}{!}{%
  \begin{tabular}{c|c|c|c}
  \hline
  Network               & All centralities (Fifteen centralities) & Global centralities (Four global centralities) & Local centralities (Eleven local centralities) \\ \hline
  Hamsterster Friends   & 1.58             & 1.48               & 0.10               \\ \hline
  Human protein (Vidal) & 1.60             & 1.55               & 0.05               \\ \hline
  CA-GrQc               & 3.85             & 3.75               & 0.10               \\ \hline
  CA-HepTh              & 16.01            & 15.80              & 0.21               \\ \hline
  Sister-Cities         & 17.82            & 17.66              & 0.16               \\ \hline
  CA-CondMat            & 108.44           & 107.60             & 0.84               \\ \hline
  \end{tabular}%
  }
  \end{table}
\par

\begin{figure}[!t]
  \centering
  \subfloat[]{\includegraphics[width=3in]{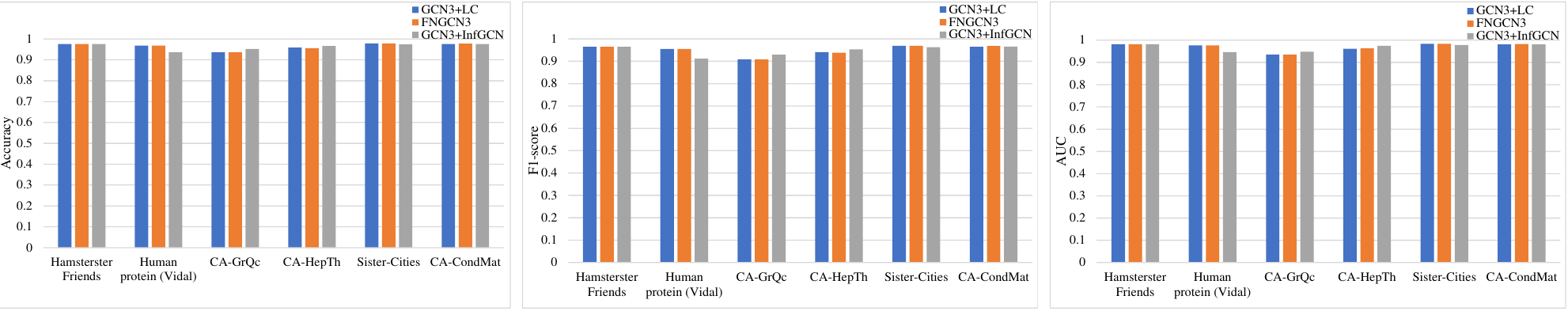}%
  \label{fig_first_case}}
  \hfil
  \subfloat[]{\includegraphics[width=3in]{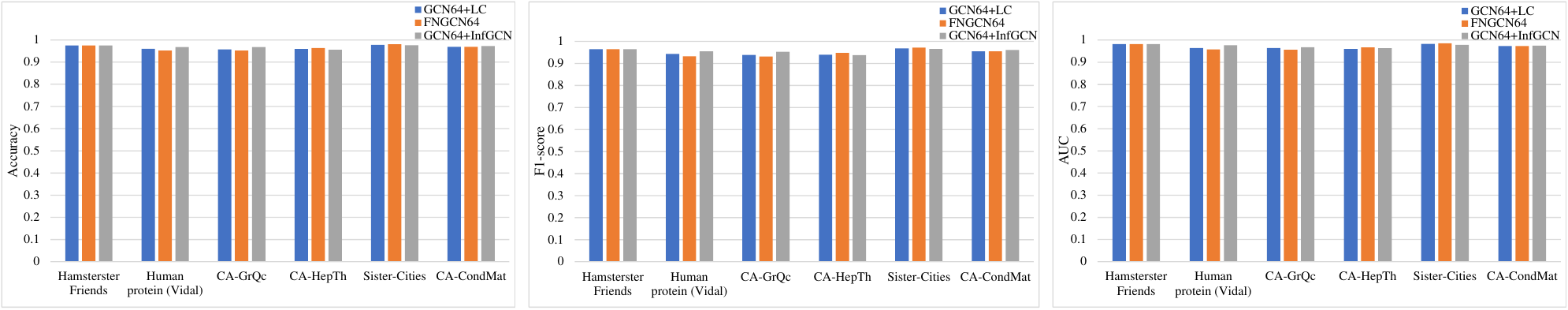}%
  \label{fig_second_case}}
  \caption{The results of the second ablation experiment, i.e., comparing FNGCN with GCN with the inputs of InfGCN and all local centralities. (a) Results obtained by  FNGCN3, GCN3 with the inputs of InfGCN(abbreviated as GCN3+InfGCN) and GCN3 with all local centralities (abbreviated as GCN3+LC, (b) Results obtained by  FNGCN64, GCN64 with the inputs of InfGCN(abbreviated as GCN64+InfGCN) and GCN64 with all local centralities (abbreviated as GCN64+LC)}
  \label{fig_sim}
  \end{figure}

\noindent \textbf{The effect of the centrality selection.} The results of the second ablation experiment are shown in Fig.5, the results of FNGCN3, GCN3 with inputs of InfGCN, and all local centralities are shown in Fig.5 (a), the results of FNGCN64, GCN64 with inputs of InfGCN, and all local centralities are shown in Fig.5 (b). According to Fig.5(a), compared with GCN3 with all local centralities, FNGCN3 leads to a little degradation on CA-HepTh in terms of accuracy and F1-score, but performs equally on Hamsterster Friends, Human protein (Vidal), CA-GrQc and Sister-Cities, brings improvement on  CA-CondMat in terms of accuracy and F1-score, on CA-HepTh  and CA-CondMat in terms of AUC. Compared with GCN3 with inputs of InfGCN, FNGCN3 leads to a little degradation on CA-GrQc and CA-HepTh, but performs equally on Hamsterster Friends  and  brings  improvement on Human protein (Vidal), Sister-Cities and CA-CondMat. According to Fig.5 (b), compared with GCN64 with all local centralities, FNGCN64 leads to a little degradation on Human protein (Vidal) and CA-GrQc, but performs equally on Hamsterster Friends and CA-CondMat, and brings  improvement on CA-HepTh and Sister-Cities. Compared with GCN64 with inputs of InfGCN, FNGCN64 leads to a little degradation on Human protein (Vidal), CA-GrQc and CA-CondMat, but performs equally on Hamsterster Friends and brings  improvement on CA-HepTh and Sister-Cities. These results indicate that FNGCN can lead to results comparable with all local centralities and the input of InfGCN, and even bring improvements in some cases, demonsting that it is feasible to use FNGCN to identify influential nodes in the network.

\begin{figure}[!t]
  \centering
  \subfloat[]{\includegraphics[width=2.7cm]{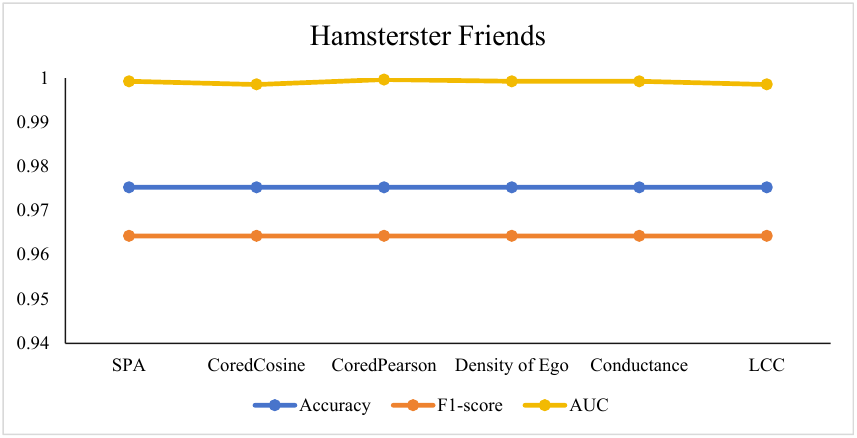}%
  \label{4.4.4-1}}
  \subfloat[]{\includegraphics[width=2.7cm]{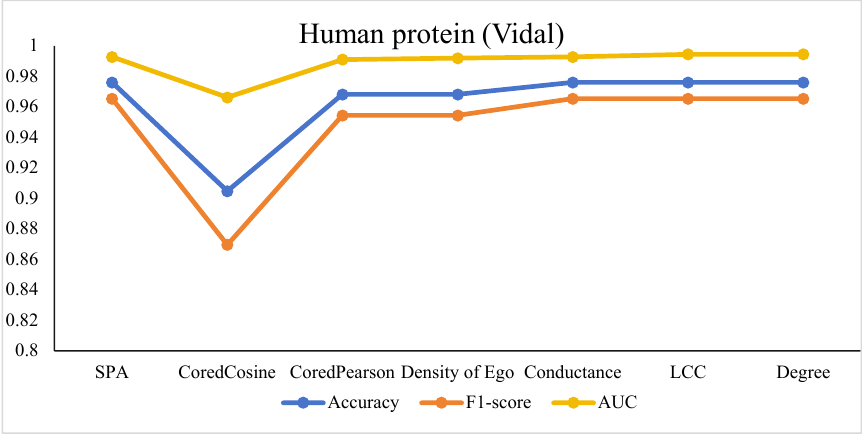}%
  \label{4.4.4-2}}
  \subfloat[]{\includegraphics[width=2.7cm]{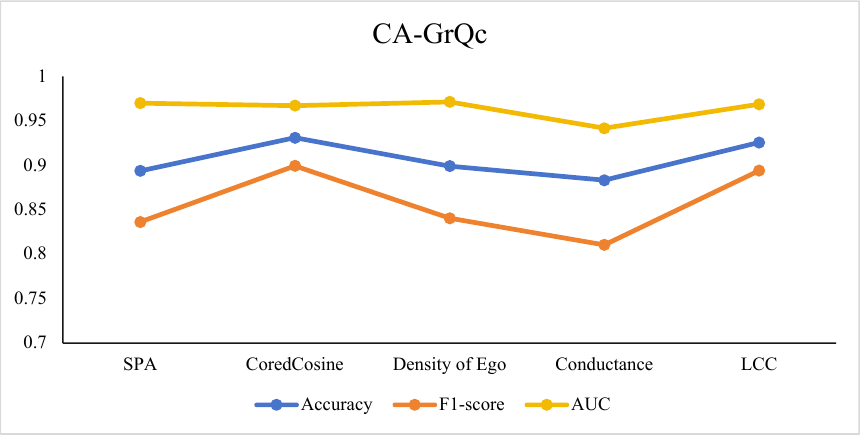}%
  \label{4.4.4-3}}
  \hfil
  \subfloat[]{\includegraphics[width=2.7cm]{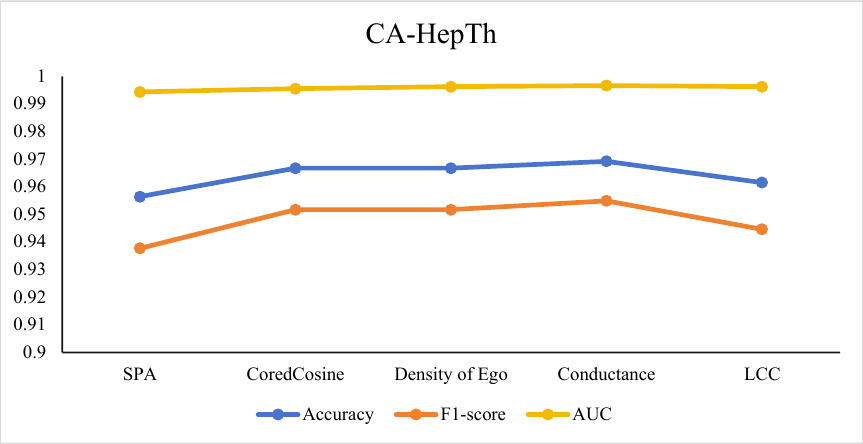}%
  \label{4.4.4-4}}
  \subfloat[]{\includegraphics[width=2.7cm]{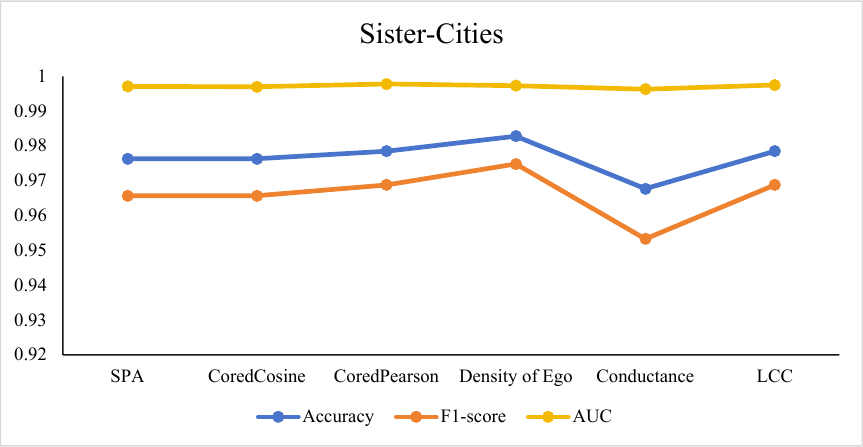}%
  \label{4.4.4-5}}
  \subfloat[]{\includegraphics[width=2.7cm]{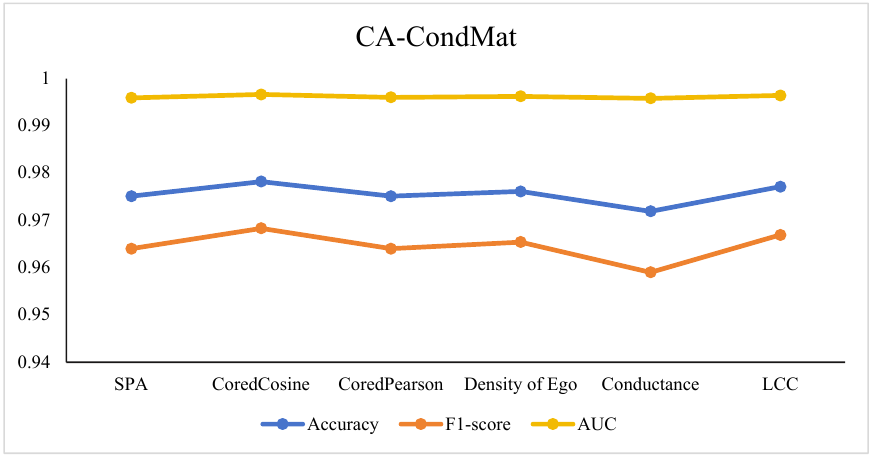}%
  \label{4.4.4-6}}
  \caption{The contribution of different local centralities based on FNGCN3}
  \label{fig_sim6}
  \end{figure}

  \textbf{The contributions of different local centralities to influential node identification.}To explore the contributions of each chosen local centrality to the task of identifying influential nodes, for each chosen local centrality, we exclude it from the node features, and evaluate the ability of the remained node features to identify influential nodes. The more the ability decreases, the greater the contribution of this local centrality has. Fig. 6 and Fig. 7 show the results regarding to the chosen local centralities based on FNGCN3 and FNGCN64, respectively. According to Fig. 6, on Hamsterster Friends, FNGCN3 produces very similar results when each of the chosen local centralities is excluded from node features. On Human protein (Vidal), FNGCN3 produces similar results regarding to all the chosen local centralities except CoredCosine, excluding which FNGCN3 produces obviously worse result. On CA-GrQc, among all the chosen local centralities, excluding Conductance of Egonet makes FNGCN produce obviously lower accuracy as well as F1-score and AUC than excluding the other local centralities; excluding SPA and Density of Egonet also results in lower accuracy and F1-score. On CA-HepTh, excluding SPA and LCC makes FNGCN3 produce lower accuracy and F1-score, while excluding the other local centralities leads to very similar results. On Sister-Cities and CA-CondMat, excluding Conductance of Egonet leads to lower accuracy and F1-score than excluding the others, and the results are similar in the other cases.
  \begin{figure}[!t]
    \centering
    \subfloat[]{\includegraphics[width=2.7cm]{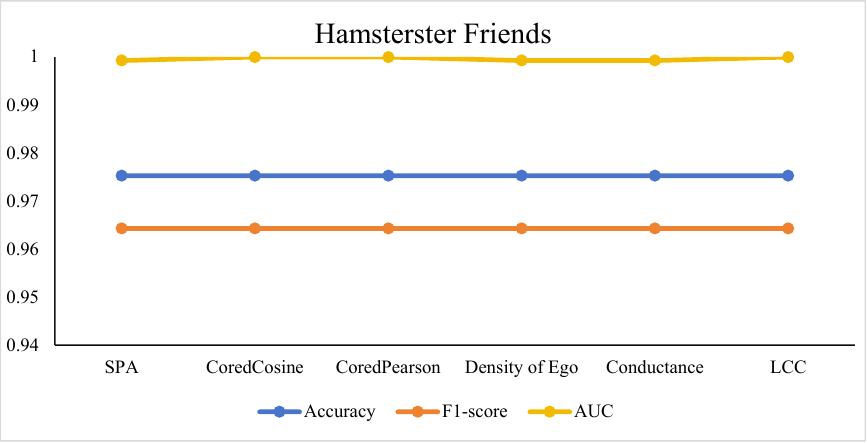}%
    \label{4.4.4-8}}
    \subfloat[]{\includegraphics[width=2.7cm]{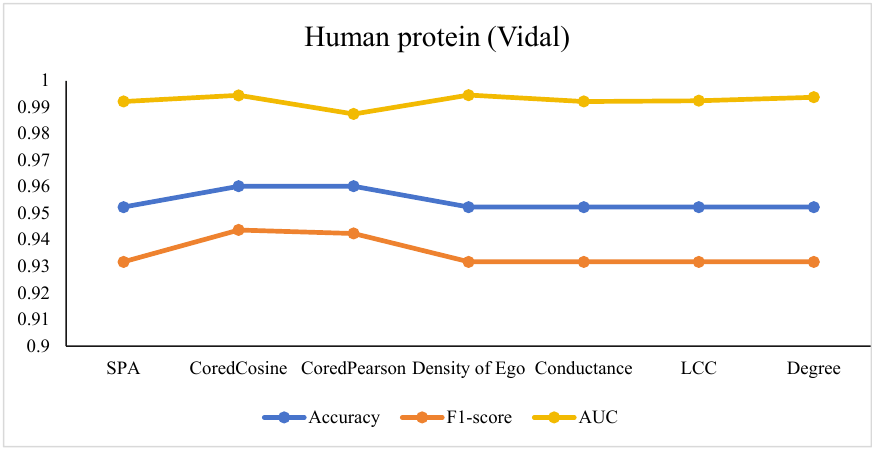}%
    \label{4.4.4-9}}
    \subfloat[]{\includegraphics[width=2.7cm]{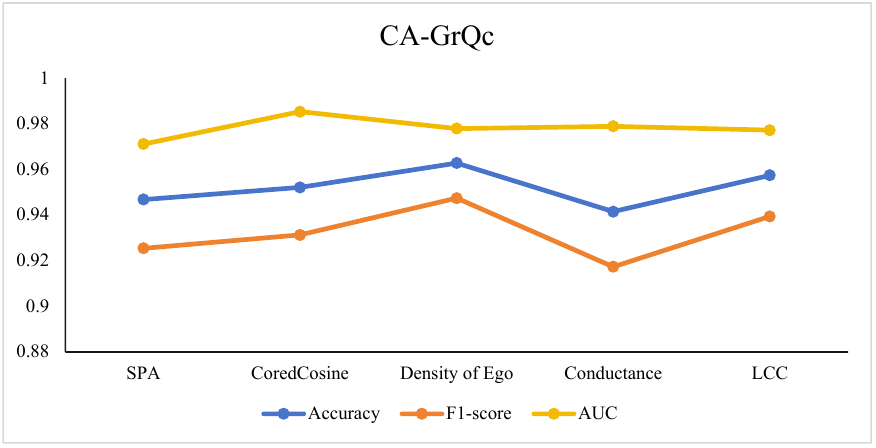}%
    \label{4.4.4-10}}
    \hfil
    \subfloat[]{\includegraphics[width=2.7cm]{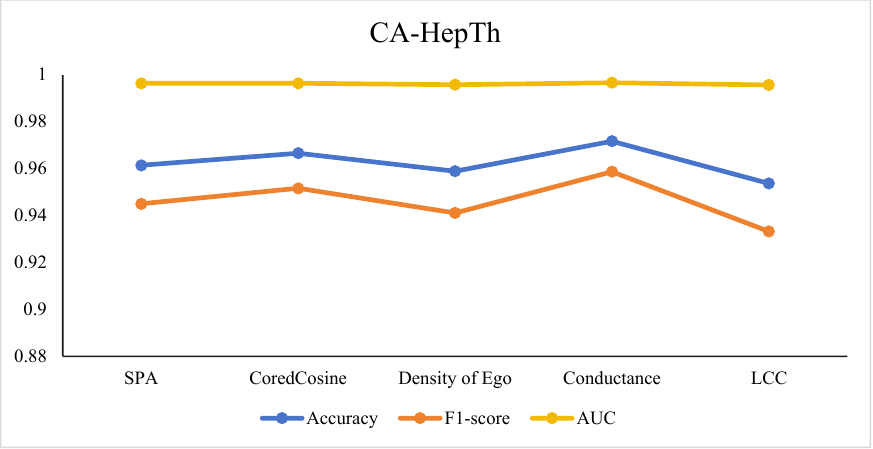}%
    \label{4.4.4-11}}
    \subfloat[]{\includegraphics[width=2.7cm]{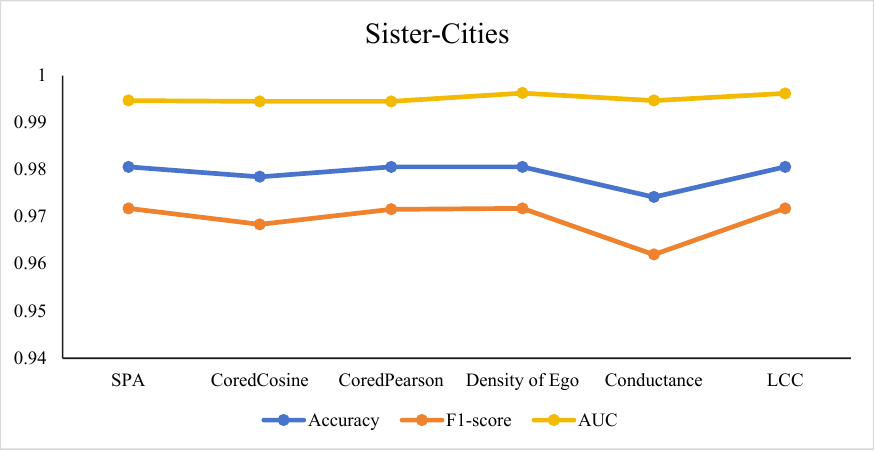}%
    \label{4.4.4-12}}
    \subfloat[]{\includegraphics[width=2.7cm]{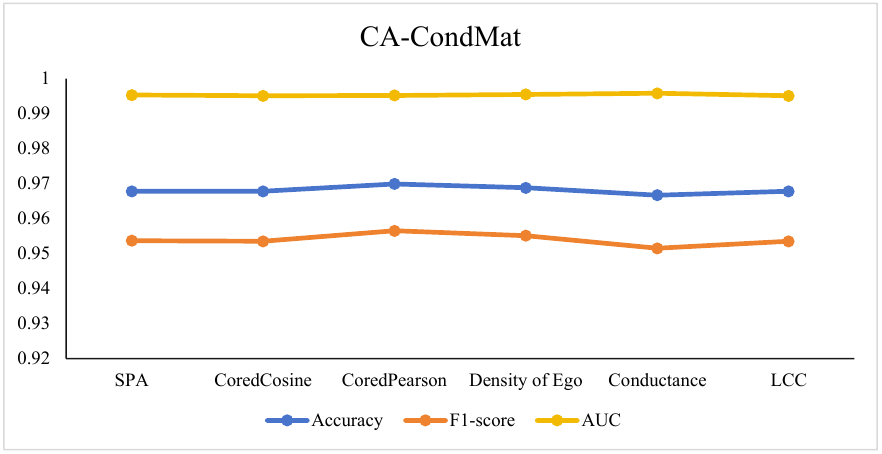}%
    \label{4.4.4-13}}
    \caption{The contribution of different local centralities based on FNGCN64}
    \label{fig_sim7}
    \end{figure}
  \par
  According to Fig.7, on Hamsterster Friends and Sister-Cities, the observation corresponding to FNGCN64 is similar to that corresponding to FNGCN3. On Human Protein (Vidal), among all the chosen local centralities, FNGCN64 produces higher accuracy and F1-score when excluding CoredCosine and CoredPearson, and produces lower AUC when excluding CoredPearson; and in the other cases, FNGCN64 performs similarly. On CA-GrQc, FNGCN64 produces the lowest accuracy and F1-score when excluding Conductance of Egonet, and the second lowest accuracy and F1-score when excluding SPA; excluding SPA also results in the lowest AUC. On CA-HepTh, excluding LCC makes FNGCN64 produce the lowest accuracy and F1-score, followed by Density of Egonet and SPA. On CA-CondMat, excluding Conductance of Egonet results in a slightly lower accuracy and F1-score than excluding the other chosen local centralities, and in the other cases FNGCN64 performs similarly.
  \par
  From the results above, it can be inferred that Conductance of Egonet is most important to FNGCNs, especially FNGCN64; SPA, LCC and CoredCosine are also important to FNGCNs on some cases, and in the other cases, the contributions of the chosen local centralities to FNGCNs are very close.

\section{Conclusion}
 \label{sec:Conclusion}
 \noindent In this paper, we study the problem of identifying influential nodes in complex networks. In previous works, the traditional methods use one or several centrality metrics to evaluate the influence of nodes, and identify the nodes with the best centrality values as influential nodes, and identify the nodes with the best centrality values as influential nodes, leading to one-sided results; traditional machine learning and deep learning based methods directly utilize several common centralities (most of them are global centralities) as node features to distinguish influential nodes from non-influential nodes, and ignore many local centralities that are much more efficient, leading to inaccurate results in some cases as well as costing much more time. To solve these issues, we propose FNGCN to identify influential nodes, which turns to local centralities and selects the most suitable local centralities as node features by exploring the complex relationships between a numbers of local centralities from the perspective of network analysis. We further explore the impact of the number of hidden GCN layers on FNGCN, and the contribution of each node feature to FNGCN. Comprehensive experiments across various datasets show that: 1) FNGCN outperforms the compared methods including state-of-the-art models overall, and FNGCN cost much less time to obtain node features than those with global centralities; 2) deep GCNs performs better than shallow GCNs in some cases, but the impact of the number of hidden GCN layers on FNGCN is slight; 3) local centralities, which are much efficient than global centralities, are sufficient as node features to identify influential nodes; but it is necessary to select suitable local centralities, according to the relationships between them, to prepare node features; 4) the centrality of Conductance of Egonet is most important to FNGCN. However, FNGCN requires constructing feature network, which is used to represent and analyse the relationships between different local centralities, for each network. In addition, the proposed approach is now only tested on one type of GCN, but various types of GNNs have been proposed in the literature. In the future work, in addition to making effort on the parallel implementation of FNGCN to efficiently handle large-scale networks, it is worth exploring a large number of datasets from various domains to obtain pervasive relationship patterns of local centralities, at least for the same type of networks. It is also worth trying different types of GNN to explore the generalization ability of our approach, and adapting FNGCN to handle dynamic networks in the future work.
 \newline
 \newline
 \textbf{Acknowledgements} This work is supported by Natural Science Foundation of China (No. 61802034), Sichuan Science and Technology Programs(Nos. 2022YFQ0017).
 \newline
 \newline
 \textbf{Author contributions} Siyuan Yin: Conceptualization, Methodology, Formal analysis, Data Curation, Writing - original draft preparation. Yanmei Hu: Conceptualization, Methodology, Formal analysis and investigation, Writing - original draft preparation. Yihang Wu: Writing - review and editing, and supervision. Xue Yue: Formal analysis, Writing - review and editing, and supervision.  Yue Liu: Writing - review and editing, and supervision.
 \newline
 \newline
 \textbf{Funding} This work is supported by Natural Science Foundation of China (No. 61802034), Sichuan Science and Technology Programs(Nos. 2022YFQ0017).
 \newline
 \newline
 \textbf{Data availability statement} The datasets generated during and/or analysed duringthe current study are available from the corresponding author on reasonablerequest.
 \section*{Declarations}
 \noindent\textbf{Conflict of interest} The authors declare that they have no conflict of interest.
 \newline
 \newline
 \textbf{Ethics approval} This article does not contain any studies with human.
 \newline
 \newline
 \textbf{Human and animal rights} No violation of human and animal rights is
 involved.

\bibliographystyle{IEEEtran}

\bibliography{IEEEabrv,mylib}



\end{document}